\documentclass[prd,aps,showpacs,preprintnumbers,amssymb]{revtex4}
\usepackage{graphicx}% Include figure files

\begin{document}

\title{%
\hfill{\normalsize\vbox{%
\hbox{\rm SU-4252-909}
 }}\\
{Doubly Perturbed $S_3$ neutrinos and
the $s_{13}$ mixing parameter}}

\author{Renata Jora
$^{\it \bf a}$~\footnote[1]{Email:
 rjora@ifae.es}}

\author{Joseph Schechter
 $^{\it \bf b}$~\footnote[2]{Email:
 schechte@phy.syr.edu}}

\author{M. Naeem Shahid
$^{\it \bf b}$~\footnote[3]{Email:
   mnshahid@phy.syr.edu            }}

\affiliation{$^{\bf a}$ Grup de Fisica Teorica,
Universitat Autonoma de Barcelona, E -08193
Belaterra (Barcelona), Spain}
 
\affiliation{$^ {\bf \it b}$ Department of Physics,
 Syracuse University, Syracuse, NY 13244-1130, USA,}

\date{\today}

\begin{abstract}
We further study a predictive model for the masses 
and mixing matrix of three Majorana neutrinos. At zeroth order
the model yielded degenerate neutrinos and a generalized 
``tribimaximal" mixing matrix. At first order the mass
 splitting was incorporated and the tribimaximal mixing
matrix emerged with very small corrections but with a zero value
for the parameter $s_{13}$. In the present paper
a different, assumed weaker, perturbation is included which
gives a non zero value for $s_{13}$ and further corrections to
other quantities. These corrections are worked out
and their consequences discussed under the simplifying 
assumption that
 the conventional CP violation phase vanishes. It is shown that the
existing measurements of the parameter $s_{23}$ provide strong
bounds on $s_{13}$ in this model.

\end{abstract}


\maketitle

\section{Introduction}

    At present, the particle physics community is planning, as a 
follow-up to the enormously important experiments of the last decade
\cite{kamland}-\cite{minos}, an 
extensive program with the goal of more accurately
understanding the neutrino masses and mixings. There is really no accepted 
theory for an a priori prediction of these quantities. Hence it seems
worthwhile to investigate in detail various theoretical models to develop
plausible scenarios which might be tested. 

    Here we look more closely at  a particular model presented in 
\cite{jns} and
further studied in \cite{cw} and in \cite{jss}. That model assumed an 
initial permutation symmetry ($S_3$) which is motivated by the fact that
the 3 $\times$ 3  matrix which transforms the defining representation  
to irreducible form is, up to a single parameter rotation,
 the same as
 the "tribimaximal" matrix, which is in, at least, rough
 agreement with the present
experimental situation. The tribimaximal form is taken to be:
\begin{equation}
    K_{TBM}=\left(%
\begin{array}{ccc}
  \frac{-2}{\sqrt{6}} & \frac{1}{\sqrt{3}} & 0 \\
  \frac{1}{\sqrt{6}} & \frac{1}{\sqrt{3}} & \frac{1}{\sqrt{2}} \\
  \frac{1}{\sqrt{6}} & \frac{1}{\sqrt{3}} & \frac{-1}{\sqrt{2}} \\
\end{array}%
\right) \equiv R.
\label{R}
\end{equation}
 The assumption was also made that, at zeroth 
order, the three neutrinos are degenerate. It may be seen
 from Table I of \cite{jns} that this is plausible for  
 a large range of possible fits to the data. However such an assumption
at first seems inconsistent with permutation symmetry which suggests two 
of the three neutrinos to be degenerate (however $\it{not}$
the two "solar" neutrinos) and different 
in mass from the 
third. The proposed solution to this problem called for the introduction 
of a Majorana type phase, which does not affect the usual neutrino 
oscillations but does affect the rate for neutrinoless double beta decay.
The complications involved in obtaining a suitable Higgs scheme for 
both the neutrino mass matrix and the charged lepton mass matrix
(which can be arranged to be proportional to the unit matrix) in this 
approach are discussed in some detail in \cite{jns}.

    Of course, many interesting 
 different models for neutrinos 
based on 
permutation symmetry have been discussed for a long time \cite{w}-
\cite{dgr}. In addition, many interesting models with
 similar approaches to 
the tribimaximal mixing matrix have been vigorously pursued \cite{fx}
-\cite{mmp}.

     In the model under present discussion, the zeroth order piece 
of the prediagonal Majorana neutrino mass matrix has the
well known $S_3$ invariant form:
\begin{equation}
M_\nu=\alpha
\left[
\begin{array}{ccc}
1&0&0\\
0&1&0\\
0&0&1
\end{array}
\right]+\beta
\left[
\begin{array}{ccc}
1&1&1\\
1&1&1\\
1&1&1\\
\end{array}
\right] \equiv \alpha {\bf 1}+\beta d .
\label{zero}
\end{equation}
Here $\alpha$ and $\beta$ are, in general, complex numbers
 while $d$ is
usually called the ``democratic" matrix.
As discussed in detail in \cite{jss} and \cite{jns}, we take 
\begin{equation}
\alpha=-i|\alpha|e^{-i\psi/2},
\label{alpha}
\end{equation}
where the physical phase $\psi$ lies in the range:
\begin{equation}
0<\psi\leq \pi.
\label{psirange}
\end{equation}
For the assumed initial degeneracy,
 $|\alpha|$ is related to 
$\beta$, assumed real, by:
\begin{equation}
|\alpha|=\frac{3\beta}{2sin(\psi/2)}.
\label{alphabeta}
\end{equation}
 The two zeroth order parameters are the 
degenerate neutrino masses, $|\alpha|$ and the phase 
$\psi$ which contributes to the neutrinoless
double beta decay amplitude.

The first order perturbation treated in 
\cite{cw} and \cite{jss} is
\begin{equation}
\Delta=\left(%
\begin{array}{ccc}
  0 & 0 & 0 \\
  0 & t & u \\
  0 & u & t \\
\end{array}%
\right)
\label{23pert}
\end{equation}
where $t$ and $u$ are parameters.
In general, $t$ and $u$ may be complex but they were assumed
real for simplicity. This perturbation is well known as
the ``mu-tau" symmetry \cite{mutau}-\cite{Gutmutau}. The assumed 
zeroth 
order degeneracy (which actually may be relaxed, if desired)
forces us to use degenerate perturbation theory. Then $\Delta$
turns out (eg, section II of \cite{jss}) to be the only possible 
choice which forces the desired 
tribimaximal form (as opposed to the generalized tribimaximal form)
of the first order mixing matrix. 

    Here we will choose for the second order perturbation, the matrix:
\begin{eqnarray}
\Delta^{\prime}=\left(\begin{array}{ccc}
t^{\prime} & u^{\prime} & 0\\
u^{\prime} &t^{\prime} & 0\\
0 & 0 & 0\end{array}\right).
\label{12pert}
\end{eqnarray}
For simplicity we again consider the parameters,
$t'$ and $u'$ to be real.

    Note that this second order perturbation  
preserves the $S_2$ subgroup which involves the 1-2
interchange. One might wonder about also including a perturbation,
$\Delta''$ which preserves the 1-3 $S_2$ subgroup.
However, that is not expected to give anything new since the 
combination of Eqs.(\ref{zero}), (\ref{23pert})
 and (\ref{12pert}) already 
has the same number of parameters as the most 
general symmetric matrix, $M_\nu$.

   The combination of Eqs.(\ref{zero}), (\ref{23pert})
 and (\ref{12pert}) was motivated by the group theory treatment 
of the strong interactions before QCD which led for example to the 
Gell-Mann Okubo mass formula \cite{gmo}. In that case the initial 
term was flavor SU(3) invariant, the next term was
 invariant under the 
SU(2) isospin subgroup  while the smallest last term 
was invariant under the SU(2) ``U-spin" subgroup. In the present case 
the zeroth order term has the discrete group S$_3$ invariance and
 two different S$_2$ subgroups are left invariant by the two 
perturbations.

\section{Perturbation analysis}

    In \cite{jss} we diagonalized the needed 
symmetric matrix:

\begin{eqnarray}
&& R^T( \alpha{\bf 1} +\beta d +\Delta)R =
\nonumber \\
&& \alpha{\bf 1} +
\left(%
\begin{array}{ccc}
  t+u & \frac{\sqrt{2}}{3}(t+u) &
0\\
  \frac{\sqrt{2}}{3}(t+u) & 3\beta+ \frac{2}{3}(t+u) &
0 \\
  0 & 0 &
t-u \\
\end{array}%
\right).
\label{tobediag}
\end{eqnarray}

The diagonalization of this matrix gave the first 
order neutrino mixing matrix, $K^{(1)}$ as 
\begin{equation}
   K^{(1)}=RR_1,
\label{Kfirst}
\end{equation}

where, 

\begin{equation}
R_1 \approx
\left(%
\begin{array}{ccc}
  1 & \frac{\sqrt{2}}{9\beta}(t+u) & 0 \\
  -\frac{\sqrt{2}}{9\beta}(t+u) & 1 &
0 \\
  0 & 0 &
1 \\
\end{array}%
\right).
\label{R1}
\end{equation}
This results in a diagonalization with complex eigenvalues.
To make these real positive we multiplied $K^{(1)}$ on the right 
by a suitable diagonal matrix of phases.

To include the $2^{nd}$-order perturbation, Eq. (\ref{12pert}), 
we must diagonalize,

\begin{eqnarray}
H&=&R_1^T R^T(\alpha I + \beta d +\Delta + \Delta^\prime)RR_1
\nonumber\\
&\equiv&H^{0}+H^\prime
\label{ham}
\end{eqnarray}

where, after some computation and neglect of still higher order
terms, we obtain:

\begin{eqnarray}
H^{\prime}=R_1^T R^T \Delta^\prime RR_1 \approx
\left(\begin{array}{ccc}
\frac{5}{6}t^{\prime}-\frac{2}{3}u^{\prime} & -\frac{1}{3\sqrt{2}}
(t^{\prime}+u^{\prime})
 & \frac{1}{2\sqrt{3}}(t^{\prime}-2u^{\prime})\\
-\frac{1}{3\sqrt{2}}(t^{\prime}+u^{\prime}) & \frac{2}{3}
(t^{\prime}+u^{\prime}
) & \frac{1}{\sqrt{6}}(t^{\prime}+u^{\prime})\\
\frac{1}{2\sqrt{3}}(t^{\prime}-2u^{\prime}) & 
\frac{1}{\sqrt{6}}(t^{\prime}+u^{\prime}) &
 \frac{1}{2}t^{\prime}\end{array}\right).
\end{eqnarray}

We introduced the notation $H^0$ (Everything in Eq.(\ref{ham})
except for $\Delta^{\prime}$)
 and $H^{\prime}$ to indicate that,
rather than making an explicit diagonalization we will regard, 
the result to first order as a  ``zeroth order Hamiltonian", the
given second 
order term, Eq.(\ref{12pert}) as a ``first order
 perturbation" and use ordinary 
quantum mechanics perturbation theory to proceed. In that approach
 one has of course the 
corrections to the energies as:

\begin{equation}
E_n^{\prime}= <\psi_n\vert H^{\prime}\vert\psi_n>,
\label{qmenergy}
\end{equation}

while the corrections to the eigenvectors are,
 
\begin{equation}
\psi_{m}^{(1)}=\sum_{n\ne m} \frac{<\psi_{n}\vert 
H^{\prime}\vert\psi_{m}>}{E_{m}-E_{n}}\psi_n.
\label{correctvecs}
\end{equation}
A more general perturbation approach, which gives
the same results, is discussed in the Appendix.
The lepton mixing matrix up to and including
 second order then reads:

\begin{equation}
K=RR_1R_2P=(\psi_1,\psi_2,\psi_3)P,
\end{equation}

where the $\psi_i$ are the columns of $RR_1R_2$ and 
furthermore 
$P$ is the phase matrix needed for the neutrino masses
 to be real positive; explicitly,

\begin{eqnarray}
&& 
\psi_{1}=\frac{1}{\sqrt{6}}\left(\begin{array}{c}
-2-2\frac{t+u}{9\beta}+2\frac{t^{\prime}+u^{\prime}}{\beta}\\
 1-2\frac{t+u}{9\beta}-3\frac{t^{\prime}-2u^{\prime}}{t-2u}+\frac{t^{\prime}
+u^{\prime}}{9\beta}\\
 1-2\frac{t+u}{9\beta}+3\frac{t^{\prime}-2u^{\prime}}{t-2u}+\frac{t^{\prime}
+u^{\prime}}{9\beta}
\end{array}\right),
\nonumber \\
&& \psi_{2}=\frac{1}{\sqrt{3}}\left(\begin{array}{c}
 1-2\frac{t+u}{9\beta}+\frac{t^{\prime}+u^{\prime}}{9\beta}\\
 1+\frac{t+u}{9\beta}-\frac{t^{\prime}+u^{\prime}}{18\beta}
+\frac{t^{\prime}+
u^{\prime}}{6}\\
 1+\frac{t+u}{9\beta}-\frac{t^{\prime}+u^{\prime}}{18\beta}
-\frac{t^{\prime}+u^{\prime}}{6}\end{array}\right),
\nonumber \\
&&\psi_{3}=\frac{1}{\sqrt{2}}\left(\begin{array}{c}
-\frac{1}{2}\frac{t^{\prime}-2u^{\prime}}{t-2u}-
\frac{t^{\prime}+u^{\prime}}{9\beta}\\
1+\frac{1}{4}\frac{t^{\prime}-2u^{\prime}}{t-2u}-
\frac{t^{\prime}+u^{\prime}}{9\beta}\\
-1+\frac{1}{4}\frac{t^{\prime}-2u^{\prime}}{t-2u}-
\frac{t^{\prime}+u^{\prime}}{9\beta}\end{array}\right),
\label{evecs}
\end{eqnarray}

and the phase matrix has the form,

\begin{eqnarray}
P=\left(\begin{array}{ccc}
e^{-i \tau} & 0 & 0\\
0 & e^{-i \sigma} & 0\\
0 & 0 & e^{-i \rho}\end{array}\right),
\label{p}
\end{eqnarray}

wherein,
\begin{eqnarray}
\tau & \approx & 
\frac{\pi}{2}+\frac{1}{2}tan^{-1}[\frac{cot(\psi/2)}{1-\frac{2(t+u)}
{9\beta}-\frac{5t^{\prime}}{9\beta}-\frac{4u^{\prime}}{9\beta}}]
\nonumber \\
\sigma &\approx & \pi-\frac{1}{2}tan^{-1}[\frac{cot(\psi/2)}
{1+\frac{4(t+u)}{9\beta}
+\frac{4(t^{\prime}+u^{\prime})}{9\beta}}] \nonumber \\
\rho & \approx & 
\frac{\pi}{2}+\frac{1}{2}tan^{-1}[\frac{cot(\psi/2)}{1-\frac{2(t+u)}
{3\beta}-\frac{t^{\prime}}{3\beta}}].
\label{tsr}
\end{eqnarray}
Note that we are free to subtract $(\tau + \sigma + \rho)/3$ from each of 
these
three entries. Then the sum of the modified three entries will vanish 
in accordance with the requirement that there be only two 
independent Majorana phases.
The real positive neutrino masses to second order are then:

\begin{eqnarray}
&& 
m_{1}\approx\frac{3}{2}\beta csc\frac{\psi}{2}[1-\frac{2}
{9\beta}(t+u+\frac{5}{
2}t^{\prime}-2u^{\prime})sin^{2}\frac{\psi}{2}],
\nonumber \\
&&m_{2}\approx\frac{3}{2}\beta csc\frac{\psi}{2}[1+\frac{4}{9\beta}
(t+u+t^{\prime}
+u^{\prime})sin^{2}\frac{\psi}{2}],
\nonumber \\
&&m_{3}\approx\frac{3}{2}\beta csc\frac{\psi}{2}[1-\frac{2}
{3\beta}(t-u+\frac{1
}{2}t^{\prime})sin^{2}\frac{\psi}{2}].
\label{masses}
\end{eqnarray}

Notice that the zeroth order masses have the characteristic strength,
$\beta$ while the first order masses are suppressed by $(t,u)/\beta$
and the second order masses are suppressed by $(t^{\prime},u^{\prime})/
\beta$.

Also notice that the absolute values of the neutrino masses
depend on the Majorana phase, $\psi$. However, the
lepton number conserving neutrino oscillations can not depend 
on a Majorana phase \cite{nunubar}. As a check of this we see that the 
phase 
 $\psi$ cancels out when one considers the mass ${\it differences}$,

\begin{eqnarray}
&& 
A\equiv m_{2}^{2}-m_{1}^{2}\approx 3\beta(t+u)+\frac{9}{2}\beta 
t^{\prime},
\nonumber \\
&&B\equiv 
m_{3}^{2}-m_{2}^{2}\approx 
\beta(-5t+u)-\beta(\frac{7}{2}t^{\prime}+2u^{\prime}), \nonumber \\
&&C\equiv  
m_{3}^{2}-m_{1}^{2}\approx 
2\beta(-t+2u)+\beta(t^{\prime}-2u^{\prime}).
\label{masses}
\end{eqnarray}

    Of course, $A$, $B$ and $C$ are not independent. There are two, 
presently unresolved, experimental possibilites:

\begin{eqnarray}
&&Type 1:\hspace{1cm} m_3>m_2>m_1,
\nonumber \\
&&Type 2:\hspace{1cm} m_2>m_1>m_3.
\label{spectrumtype}
\end{eqnarray}.

The corresponding relations are:
 \begin{eqnarray}
&&Type 1:\hspace{1cm} |C| = |B| + A ,
\nonumber \\
&&Type 2:\hspace{1cm} |C| = |B| - A.
\label{abc}
\end{eqnarray}.

These relations were obtained by using the
 known positive sign of $A$ and that
only the two possibilities $m_3^2 > m_2^2 > m_1^2$
and  $m_2^2 > m_1^2 > m_3^2$ are allowed.
In the literature some works specify $A$ and $|B|$ while others
specify $A$ and $|C|$.

The following best fit values for the perturbation parameters
$\beta t$ and $\beta u$ were given 
in the first
order treatment  \cite{jss}:

\begin{eqnarray}
&&\beta t \approx -4.13 \times 10^{-4} eV^2,
\nonumber   \\
&&\beta u \approx 4.39 \times 10^{-4} eV^2, \quad
Type 1
\label{solve1}
\end{eqnarray}

\begin{eqnarray}
&&\beta t \approx 4.21 \times 10^{-4} eV^2,
\nonumber   \\
&&\beta u \approx -3.94 \times 10^{-4} eV^2 \quad 
Type 2.
\label{solve2}
\end{eqnarray}

\section{Elements of the mixing matrix}

     We employ the following parameterization \cite{mns} of the leptonic 
mixing matrix, $K$:

\begin{equation}
K=
\left(%
\begin{array}{ccc}
   c_{12} c_{13} & s_{12} c_{13}& s_{13} e^{-i\gamma} \\
  -s_{12}c_{23}-c_{12}s_{13}s_{23}e^{i\gamma}&
   c_{12}c_{23}-s_{12}s_{13}s_{23}e^{i\gamma}&
   c_{13}s_{23} \\
   s_{12}s_{23}-c_{12}s_{13}c_{23}e^{i\gamma}&
  -c_{12}s_{23}-s_{12}s_{13}c_{23}e^{i\gamma}&
   c_{13}c_{23} \\
\end{array}%
\right)P ,
\label{usuparam}
\end{equation}
where $c_{12}$ is short for $cos\theta_{12}$ for example.
$P$ is the diagonal matrix of Majorana type phases given in 
Eqs.(\ref{p}) and (\ref{tsr}) for the present model. For simplicity we are
presently neglecting the conventional CP violation and thus setting 
$\gamma=0$. To specify $s_{12}$, $s_{13}$ and $s_{23}$, it is clearly 
sufficient to compare the (1-2), (1-3) and (2-3) matrix
 elements of $K$ in 
Eq.(\ref{usuparam}) with those calculated in Eq.(\ref{evecs}). This 
yields:
\begin{eqnarray}
&&
s_{12}c_{13}=\frac{1}{\sqrt{3}}-\frac{2}{\sqrt{3}}
\frac{t+u}{9\beta}+\frac{1}{
\sqrt{3}}\frac{t^{\prime}+u^{\prime}}{9\beta}, 
\nonumber \\
&&
s_{13}=-\frac{1}{2\sqrt{2}}\frac{t^{\prime}-2u^{\prime}}{t-2u}-
\frac{1}{\sqrt{2}}\frac{t^{\prime}+u^{\prime}}{9\beta},
\nonumber \\
&& s_{23}c_{13}=\frac{1}{\sqrt{2}}+\frac{1}{4\sqrt{2}}
\frac{t^{\prime}-2u^{\prime}
}{t-2u}-\frac{1}{\sqrt{2}}\frac{t^{\prime}+u^{\prime}}{9\beta}.
\label{delang}
\end{eqnarray}

    For an initial orientation we see that at zeroth order,
 $s_{13}$ vanishes and also $K$ has the tribimaximal form. When the 
first order perturbation characterized by $t$ and $u$ is added, neither
 $s_{13}$ nor $s_{23}$ change. However  $s_{12}$ is somewhat modified as 
discussed previously in section IV of \cite{jss}.
 When the second order perturbation characterized 
by $t^{\prime}$ and $u^{\prime}$ is added, $s_{13}$ finally becomes non-zero
while both $s_{12}$ and $s_{23}$ suffer further corrections.

   But something unusual is happening; there are terms for 
$s_{13}$ and $s_{23}$ which behave like $t^{\prime}/t$ and 
are manifestly of first order in strength. These arise from 
the energy difference denominator in Eq.(\ref{correctvecs}).
Since we had to use degenerate perturbation theory at first 
order this denominator is proportional to the first
 order ``energy" corrections rather than the zeroth order energies. 
Keeping terms of actual first order in strength we find the 
interesting relation:
\begin{equation}
 s_{13}\approx -2\delta s_{23},
\label{13formula}
\end{equation}
where $\delta s_{23}$ denotes the deviation of
 $s_{23}$ from its tribimaximal value. Also the 
good approximation $c_{13}=1$ was made.

\section{Numerical estimates}

    Already, Fogli et. al. \cite{flmpr} and Schwetz et. al. 
\cite{stv} have pointed out that detailed analysis of 
existing neutrino oscillation
experiments gives some hint for non zero $s_{13}$. Thus it
seems interesting to see what predictions emerge from 
Eq.(\ref{13formula}).

    Expanding $s_{23}$ around its ``tribimaximal value"
as $s_{23}=[s_{23}]_{TBM}+\delta s_{23}$,
 one gets:
\begin{equation}
(s_{23})^2 \approx \frac{1}{2}+\sqrt{2}\delta s_{23}.
\label{s23eq}
\end{equation}
Comparing with the results of a global analysis of the 
oscillation data given in Table A1 of \cite{stv} one then identifies,
for respectively 1$\sigma$, 2$\sigma$ and 3$\sigma$ errors:
\begin{equation}
|\delta s_{23}| = 0.05,\quad 0.08, \quad 0.11.
\label{deltas23}
\end{equation}
Note that the three cases are associated with the experimental
data relating to the 2-3 type neutrino oscillations.
Using Eq.(\ref{13formula}) then leads to the corresponding predictions,
\begin{equation}
| s_{13}| < 0.025,\quad 0.040, \quad 0.055.
\label{s13bounds}
\end{equation}
It is amusing to note that these values range
 from about 1/4 to 1/2 of the 
``best
 fit" value $|s_{13}|= 0.11$, which is also 
presented in the first column of Table A1 in \cite{stv}. 
Of course, our estimates provide a test of the present 
theoretical model for neutrino parameters and have no connection
with experimental data on $|s_{13}|$.

    As discussed above, the theoretical estimate for $|s_{13}|$, is 
of characteristic first order strength, appearing as a ratio of a 
second order quantity divided by a first order quantity. Using 
Eq.(\ref{delang}) for $s_{13}$ and neglecting the term of second order 
strength we can get an estimate of the relative second to first order 
effects:
\begin{equation}
|\frac{t^{\prime}-2u^{\prime}}{t-2u}|\approx 2\sqrt{2}|s_{13}| \approx
0.071,\quad 0.11, \quad 0.16,
\label{secondordstr}
\end{equation}
wherein Eq.(\ref{s13bounds}) was used.
Evidently the second order effects seem to be suppressed by about 
$1/10$ compared to the first order effects. On the other hand, 
as seen in Eq.(\ref{masses}), the quantities $t^\prime$ and $u^\prime$ 
enter
in the true second order corrections for the neutrino mass differences. 
Thus those corrections are 
likely to be small-- on the order of ten percent of the first order
mass splittings.

\section{Summary and discussion}

    In this work, we designated the zeroth order  
parameter as $\beta$, the first order parameters as 
$t$ and $u$ and the second order parameters as
$t^\prime$ and $u^\prime$. The first order corrections
to the neutrino masses were suppressed by $(t,u)/\beta$
compared to zeroth order. For the mixing angles, the first 
order corrections had a previously obtained piece proportional
to $(t,u)/\beta$ as well as a new piece proportional to 
$(t^\prime, u^\prime)/(t,u)$. The latter term
 arose because we are using 
degenerate perturbation theory and is clearly important for 
$s_{13}$ to be non-zero and correlated to corrections 
of  $s_{23}$.

     Here, we have numerically neglected, for both masses 
and mixing angles terms proportional to 
$(t^\prime, u^\prime)/\beta$. In \cite{jss} we considered 
$(t, u)/\beta$ to be about 1/5. Here we found
a characteristic strength of $s_{13}$ to correspond to 
$(t^\prime, u^\prime)/(t,u)$ about 1/10. Both of these
magnitudes are roughly similar.

    Note that Eqs.(\ref{masses}) for the neutrino mass 
differences and 
Eqs.(\ref{delang}) for the mixing angles do contain 
pieces of actual second order strength. These should be 
interesting to study in the future when more precise data
becomes available. 

    The first order corrected formula for the
 neutrinoless double beta decay factor is given in
Eq.(51) of \cite{jss}. This was derived from Eq.(49) in
which $(s_{13})^2$ was set to zero. Now $s_{13}$ is not 
zero but its square contributes at a higher order. 
Furthermore, it is easy to see, using Eqs.(\ref{evecs}),
that the first two terms in Eq.(49)
do not have any contributions of first order
strength like 
 $(t^\prime,u^\prime)/(t,u)$.
Hence that
formula for $m_{ee}$ still holds to first order.

\section{Acknowledgments}

 We are happy to thank Amir Fariborz, Salah Nasri and Francesco Sannino
 for helpful
    discussions and encouragment. The work of R. Jora has been supported
by CICYT-FEDEF-FPA 2008-01430.
    The work of J. Schechter and M.N. Shahid was supported in part by
    the US DOE under Contract No. DE-FG-02-85ER 40231; they would also 
like
    to thank
    the CP$^3$-Origins group at the University of Southern Denmark,
where this work was started for
 their warm hospitality and partial support.

\appendix
\section{Alternative perturbation method}

We present here an alternative approach which leads to results
 in perturbation theory order
by order. This can be applied to the case at hand or more
 generally when the mass matrix is invariant at zeroth order
 under a finite group $G_0$ and then we add perturbations
 of decreasing importance in the small parameter x such
 that for example the $n^{th}$ perturbation is of
 order $x^n$ and is invariant under a smaller group $G_n$.
 The mass matrix can then be written as an expansion in x,
 \begin{equation}
 M(x)=M_0+xM_1+x^2M_2+...
 \label{exp45}
 \end{equation}
 where $M_0$ is invariant under $G_0$, $M_1$ under $G_1$
  and so on.

The eigenvalues (diagonal) and eigenvector matrices
 can also be expanded as,
\begin{eqnarray}
&&M_d(x)=M_{d0}+xM_{d1}+x^2M_{d2}+...
\nonumber\\
&&R(x)=R_0+xR_1+x^2R_2+...
\label{eig45}
\end{eqnarray}

where,
\begin{equation}
R^T(x)M(x)R(x)=M_d(x)
\label{eig567}
\end{equation}
is the eigenvalue equation.

If we differentiate Eq (\ref{eig567}) once we obtain:
\begin{eqnarray}
R^{T\prime}MR+R^TM'R+R^TMR'=M_d'
\label{firstder4356}
\end{eqnarray}

which can be written as:
\begin{eqnarray}
[M_d,R^TR']+R^TM'R=M_d'
\label{first43567}
\end{eqnarray}

Here we used the orthonormality condition
 for the eigenvector matrix:
\begin{equation}
R^{T\prime}R+R^TR'=0
\label{ort56478}
\end{equation}

Note that the matrix $R^{T\prime}R$ which appears
 in what follows is antisymmetric (in each order of
 perturbation theory) and in consequence all of its derivatives
 will be antisymmetric.

The second derivative and third derivative equations will read:
\begin{eqnarray}
&&[M_d',R^TR']+[M_d,(R^TR')']+[R^TM'R,R^TR']+
R^TM^{\prime \prime}R=M_d^{\prime\prime},
\nonumber\\
&&[M_d^{\prime},R^TR']+2[M_d',(R^TR')']+[M_d,(R^TR')^{\prime\prime}]+
\nonumber\\
&&[[R^TM'R,R^TR'],R^TR']+2[R^TM^{\prime\prime}R,R^TR']+[R^TM'R,(R^TR')']
+R^TM^{\prime\prime\prime}R=M_d^{\prime\prime\prime}
\label{third645378}
\end{eqnarray}

All commutators of diagonal matrices give zero on
 diagonal and in consequence the mass eigenvalues
are obtained from the rest of the terms.

It is clear that by setting $x=0$ one can associate the
 first derivative with the first order perturbation theory,
 second with second order and so on.
The mass eigenvalues and the matrix $R^TR'$ can be
 extracted in each order from equations like
Eq (\ref{first43567}) and Eq(\ref{third645378}).

Then one should use the orthonormality condition to
obtain the eigenvector matrix according to:
\begin{eqnarray}
&&R^T(x)R'(x)=
R_0^TR_1+x(R_1^TR_1+2R_0^TR_2)+....
\label{eigv456}
\end{eqnarray}.

Using this method and $G_0=S_3$, $G_1=S_{23}$ and
 $G_2=S_{12}$ one retrieves the eigenvalues and eigenvectors
in each order of perturbation theory. The results agree
 with those presented in the main text.

\end{document}